\documentclass[pre,twocolumn,showpacs,preprintnumbers,amsmath,amssymb]{revtex4-1}

\pdfoutput=1

\usepackage{graphicx}
\usepackage{dcolumn}
\usepackage{bm}

\begin{document}

\newcommand{\bea}{\begin{eqnarray}}
\newcommand{\eea}{  \end{eqnarray}}
\newcommand{\bit}{\begin{itemize}}
\newcommand{\eit}{  \end{itemize}}

\newcommand{\be}{\begin{equation}}
\newcommand{\ee}{\end{equation}}
\newcommand{\ra}{\rangle}
\newcommand{\la}{\langle}
\newcommand{\U}{\widetilde{U}}


\def\bra#1{{\langle#1|}}
\def\ket#1{{|#1\rangle}}
\def\bracket#1#2{{\langle#1|#2\rangle}}
\def\inner#1#2{{\langle#1|#2\rangle}}
\def\expect#1{{\langle#1\rangle}}
\def\e{{\rm e}}
\def\proj{{\hat{\cal P}}}
\def\tr{{\rm Tr}}
\def\H{{\hat H}}
\def\Hdag{{\hat H}^\dagger}
\def\Lop{{\cal L}}
\def\Ehat{{\hat E}}
\def\Edag{{\hat E}^\dagger}
\def\Shat{\hat{S}}
\def\Sdag{{\hat S}^\dagger}
\def\Ahat{{\hat A}}
\def\Adag{{\hat A}^\dagger}
\def\U{{\hat U}}
\def\Udag{{\hat U}^\dagger}
\def\Zhat{{\hat Z}}
\def\Phat{{\hat P}}
\def\Op{{\hat O}}
\def\id{{\hat I}}
\def\x{{\hat x}}
\def\P{{\hat P}}
\def\Px{\proj_x}
\def\Pr{\proj_{R}}
\def\Pl{\proj_{L}}


\title{3D classical and quantum stable structures of dissipative systems}

\author{Gabriel G. Carlo}
\email{carlo@tandar.cnea.gov.ar}
\author{Leonardo Ermann}
\email{ermann@tandar.cnea.gov.ar}
\author{Alejandro M. F. Rivas}
\email{rivas@tandar.cnea.gov.ar}
\affiliation{CONICET, Departamento de F\'\i sica, CNEA, Libertador 8250,
(C1429BNP) Buenos Aires, Argentina}
\author{Mar\'\i a E. Spina}
\affiliation{Departamento de F\'\i sica, CNEA, Libertador 8250,
(C1429BNP) Buenos Aires, Argentina}
\email{spina@tandar.cnea.gov.ar}

\date{\today}


\begin{abstract}

We study the properties of classical and quantum stable structures in a 3D parameter space corresponding 
to the dissipative kicked top. This is a model system in quantum and classical chaos 
that gives a starting point for many body examples. We are able to identify the influence of these 
structures in the spectra and eigenstates of the corresponding (super)operators. This provides 
with a complementary view with respect to the typical 2D parameter space systems found in the literature. 
Many properties of the eigenstates, like its localization behaviour can be generalized to this higher dimensional 
parameter space and spherical phase space topology. Moreover we find a 3D phenomenon --generalizable to more dimensions-- 
that we call the {\em coalescence-separation} of (q)ISSs, whose main consequence is a marked enhancement of 
quantum localization. This could be of relevance for systems which have attracted a lot of attention  
very recently. 

\end{abstract}

\maketitle

\section{Introduction}
\label{sec1}

Dissipative systems play a central role in many areas of physics. From the classical side 
the discovery of the so called isoperiodic stable structures (ISSs) in the 2D parameter 
space of the H\'enon map \cite{Gallas1} provided with a new perspective for bifurcation 
phenomena and stability properties. This important advance led to a vast amount of work. 
One of the many possible applications is on directed transport, 
where quantum dissipative ratchets have been proposed \cite{Carlo1}. This suggested  
the exploration of the quantum counterparts of the ISSs (qISSs) \cite{Carlo2} and 
revealed many general quantum to classical correspondence properties. These results 
have a wide range of applicability like for example in recent aspects of 
superconducting qubits \cite{Neill}, cold atoms \cite{Swingle}, and Bose-Einstein 
condensates \cite{Vorberg} experiments. 

On the other hand, open many body systems have received a lot of attention very recently. 
The case of the rocked open Bose-Hubbard dimer has shown the correspondence between the
interactions and bifurcations in the mean-field dynamics \cite{Hartmann}. An important 
derivation of this is the study of quantum bifurcation diagrams \cite{Ivanchenko,Yusipov}. Also, 
there is a renovated interest in the parameter space properties of classical dissipative maps 
whose complexity increases due to coupling \cite{Manchein1}. This has direct consequences in 
optimizing ratchet currents that can be affected by temperature effects \cite{daSilva}. 
Finally, the study of discrete time crystals poses new questions that could be answered by means 
of an open quantum systems perspective \cite{Gambetta}. 
All these developments motivate the study of more complicated parameter and phase spaces in 
order to verify the validity of previous results in this context and to discover new properties. 

By means of analyzing paradigmatic classical and quantum chaos models like 
the (modified) kicked rotator map, the dissipative standard map, 
and a periodically driven flux there have been many recent advances  
in our knowledge about the properties of the corresponding superoperators 
\cite{Carlo2,Carlo3,Beims}. We have elucidated the fundamental role played by ISSs and 
qISSs. In fact, the invariant states that belong to qISSs have the simple shape of
the limit cycles of ISSs only for exceptionally large regular structures. In the
majority of the cases these invariants look approximately the same as the
quantum chaotic attractors that are at the vicinity of the corresponding ISS in 
the classical parameter space. Moreover, we have proven that the 
sharp classical borders of these latter become blurred at the quantum level, 
and neighboring areas influence each other 
through quantum fluctuations (parametric tunneling). Also, 
the leading eigenstates which rule the transitory behaviour have a phase space
structure dominated by limit cycles of neighbouring ISSs, and their eigenvalues have the same periodicity. 
This leads to scarring (localization) \cite{scarring} on the corresponding unstable 
periodic orbits \cite{Carlo4}. 

In this work we study the properties of the quantum and classical 3D parameter space 
of the dissipative kicked top, which also allows us to investigate a spherical phase 
space. This is a paradigmatic model that has recently been used to study 
quantum correlations as probes of chaos \cite{Madhok}, quantum to classical 
correspondence in the vicinity of periodic orbits \cite{Kumari} (which could 
be extended to the dissipative case), and that has also served as a starting point 
for many body models \cite{Akila}. By using some of the tools developed for 2 parameter 
systems we are able to characterize the morphology of the 3D (q)ISSs. We find that 
some properties of the eigenvalues and eigenstates of the quantum superoperator are 
still valid in this case, giving them a more generic nature. The most prominent example 
is the localization behaviour of the eigenstates. Moreover, we study the 
{\em coalescence-separation} phenomenon present when having more than 2 parameters. 
The main quantum consequence is an enhancement of localization that could be of 
relevance for the many areas of research previously mentioned.

This paper is organized as follows, in Sec. \ref{sec2} we explain the details of the 
dissipative kicked top, together with some of the techniques used to study it. 
In Sec. \ref{sec3} the results that allow to characterize the 3D (q)ISSs are presented. 
In Sec. \ref{sec4} we give our conclusions.

\section{The dissipative kicked top}
\label{sec2}

The quantum map for the dissipative kicked top has the form:
\begin{equation}
\rho' =D_{\tau} \ F_J  \rho F_J^{\dag} \equiv \$ \rho,
 \label{super}
\end{equation}
where $F_J$  generates the unitary dynamics and $ D_{\tau} $ is
the dissipation propagator obtained from the integration of a
master equation for the density matrix.
The Floquet operator $F_J$ is given by:
\begin{equation}
F_J=\exp[{-i (k/2J) J_z^2}]  \exp[{-i \beta J_y}],
\label{floquet}
\end{equation}
where $ J_i $ are the components of the angular momentum $ {\bf J}
$. $k$ is the torsion parameter and $\beta$ the rotation
parameter associated with the periodic kicking of the angular
momentum ($\hbar=1$) \cite{Haakebook}.
Dissipation is modeled by the following Lindblad equation:
\begin{equation}
{d \over d t} \rho(t) = \gamma  \{ [J_-,\rho(t) J_+] +
[J_-\rho(t), J_+]\} \equiv \ \Lambda \rho(t),
 \label{lindblad}
\end{equation}
where $ J_{\pm} $ are the usual raising and lowering operators and
$\gamma$  the dissipation rate.  A dimensionless parameter $ \tau
= 2 J \gamma t $ which gives the relaxation time between two
actions of the unitary operator and thus fixes the strength of the
dissipation can be introduced  \cite{Braunbook}. In
Ref.\cite{propagator} Eq.(\ref{lindblad}) has been integrated in 
the semiclassical limit. The detailed form of the matrix
elements of  $ D_\tau = \exp ( \Lambda  \tau) $ is given in Eq.(4.6) 
of Ref.\cite{propagator}. The approximation based on a
saddle-point evaluation of the inverse Laplace transformation is
valid in a wide range of quantum numbers and propagation times,
with an error of order $ 1/J^2$.
The superoperator $ \$ $ in Eq.(\ref{super}) conserves $ J^2=j(j+1)$
and has dimension $ (2j+1)^2 \times (2j+1)^2 $. It will be
diagonalized in the basis ${\ket {jm}}$ of eigenstates of $J_z$ with
$m= -j,....,j $. The diagonalization of the
quantum $e^{\Lambda}$, is worked out by using the Arnoldi method \cite{Arnoldi}.

In the  classical limit corresponding to $ j \rightarrow \infty $
the phase space is the surface of the unit sphere, with $ \mu =
\cos \theta $ and $\phi$ as canonical variables, defining the
orientation of angular momentum  {\bf J}. The detailed expressions
defining the classical map taking $ (\mu, \phi) \rightarrow (\mu',
\phi')$ are given in Appendix A of Ref. \cite{classical}.
It consists of a rotation of the angular momentum by an angle $
\beta $ around the y -axis : 
\bea
\nonumber
\mu'&=& \mu \cos{\beta} - \sqrt{1 - \mu^2} \sin{\beta} \cos{\phi},\\ 
\nonumber
\phi'&=& ( \arcsin \left(\sqrt {\frac {1-\mu^2} {1- \mu'^2}} \sin{\phi} \right)
\theta(x') + ({\rm sign}(\phi) \pi - \\
\nonumber
    & & \arcsin \left( \sqrt {\frac {1-\mu^2}
{1- \mu'^2}} \sin{\phi} \right) \theta(-x') ) \: \mod{2\pi},\\
x'&=& \sqrt{1 -\mu^2} \cos{\phi} \cos{\beta} + \mu \sin{\beta}. 
\label{classmap}
\eea
followed by a  torsion around the z-axis :
\bea
\nonumber
\mu'&=&\mu,\\
\phi'&=& (\phi + k \mu) \mod 2 \pi. 
\eea
In Eq. \ref{classmap}  $x'$ is the
$x$ component of the angular momentum after rotation, $\theta(x)$ is 
the Heaviside theta-function and ${\rm sign}(x)$ denotes the sign
function. \
Finally the dissipative part is given by:
\bea 
\nonumber
\mu'&=&\frac{\mu - \tanh \tau} {1 - \mu \tanh \tau},\\
\phi'&=&\phi.
\eea
In order to perform the classical evolution we directly use this
map and obtain the asymptotic distributions which we use to compare 
with some properties of the quantum ones. 

\begin{figure}[t]
 \includegraphics[width=0.47\textwidth]{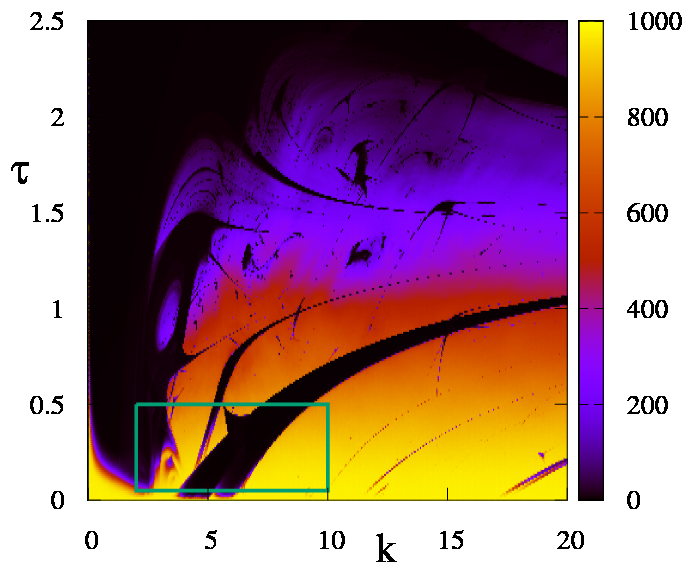}
\caption{(Color online) Participation ratio $\eta$ in parameter space $(k,\tau)$ 
for $\beta=2$. The (green) light gray rectangle represents the window of parameters 
we focus in (see Fig. \ref{fig2} d) for a better resolution).}
 \label{fig1}
\end{figure}

We have chosen to measure the chaoticity or simplicity 
of the eigenstates by means of the participation ratio $\eta=(\sum_iP(m)^2)^{-1}$, 
where $P(m)$ is the probability of $m$. This gives the number 
of basis elements that expand the quantum state. 
We generalize this concept for the classical distributions by calculating $\eta$, 
with $P(m)$ replaced by $P(\mu)$, which is a discretized limiting angular momentum 
($z$ component) distribution. This distribution is obtained after
evolving $1000$ time steps a bunch of $10000$  uniformly 
distributed random initial conditions on $(\mu, \phi)$. 
We have taken a number of $1000$ bins, which will give 
enough resolution compared to the quantum cases considered.

\section{Properties of 3D stable structures in parameter space}
\label{sec3}

We begin our study of the dissipative kicked top by exploring the classical 
parameter space. The results for a cut of this 3D space (given by $k,\beta, {\rm and} \tau$) 
at $\beta=2$ are shown in Fig. \ref{fig1}. 
The first thing we notice is that this system has the same richness as the 
2 parameter dissipative kicked rotor \cite{Carlo4}, i.e. 
we find a big regular region together with large ISSs intertwined with smaller 
shrimp-like ones, and all of them embedded in a chaotic background. 
We observe the largest regular domain (black) 
at low $k$ and large $\tau$. The second largest regular domain is a much 
smaller ISS and lies at lower $\tau$ values. From now on we focus our attention in 
the region of the parameter space where these two domains are in close 
proximity, which we have highlighted by means of a green (light gray) rectangle. 

\begin{figure}[t]
 \includegraphics[width=0.47\textwidth]{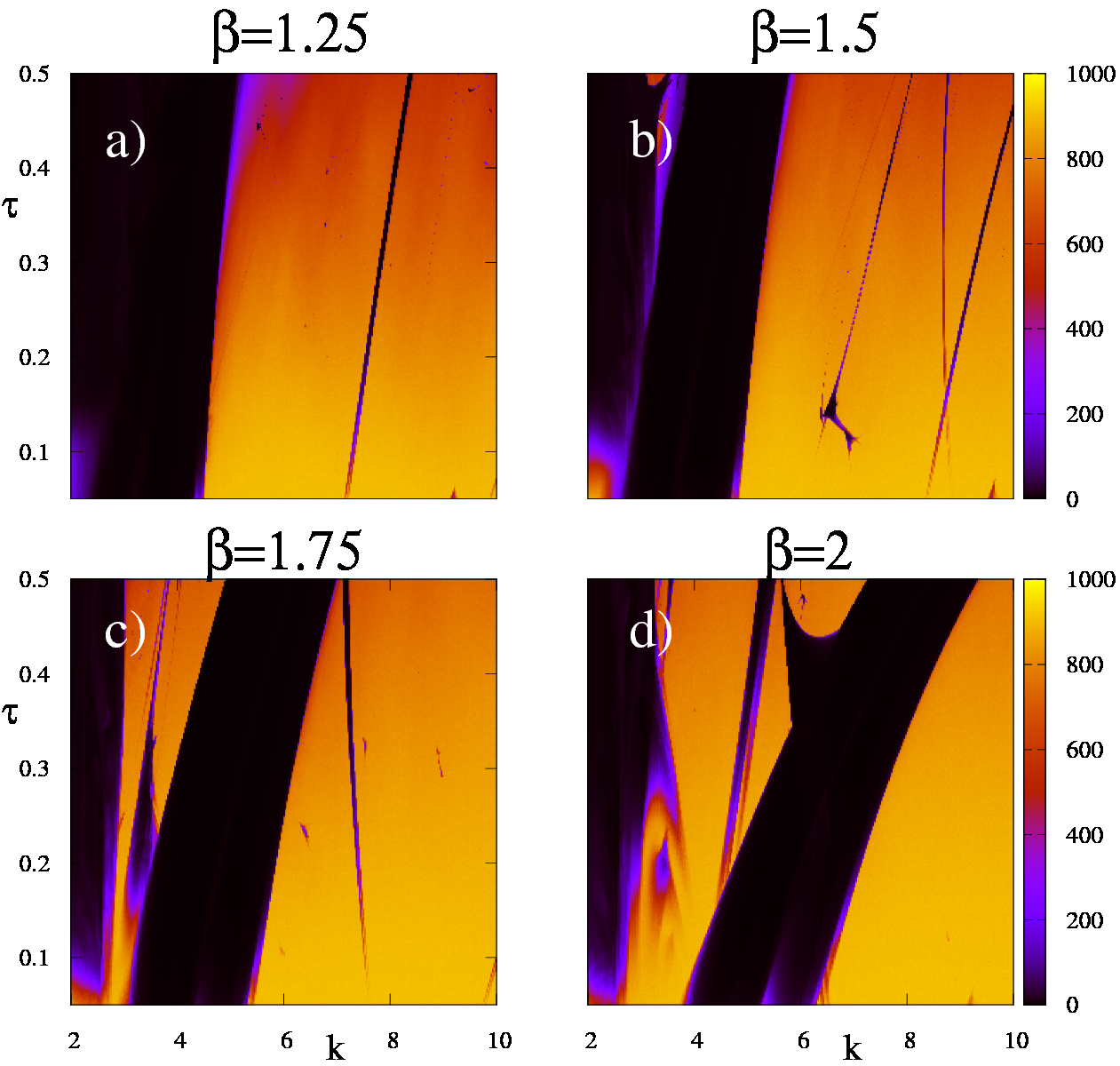}
\caption{(Color online) Participation ratio $\eta$ in parameter space $(k,\tau)$ 
for $\beta=1.25$ and $\beta=1.5$ (top left (a)) and right (b)) panels) and $\beta=1.75$ and $\beta=2$ 
(bottom left (c)) and right (d)) panels).}
 \label{fig2}
\end{figure}

This is precisely the range of $k$ and $\tau$ shown in Fig. \ref{fig2} for different 
values of $\beta$. Each panel corresponds to a screenshot of the video (linear color scale) 
included in the Supplemental 
Material that gives a better feeling of the involved 3D ISSs that build up the dissipative 
kicked top parameter space. We also show the logarithmic version of this video which provides 
more details regarding the internal structure of the ISSs. In Fig. \ref{fig2} a) the case $\beta=1.25$ 
presents just one large regular region. At $\beta=1.5$, shown in Fig. \ref{fig2} b), the large 
ISSs corresponding to the second largest regular domain is separated from the largest one. 
In Fig. \ref{fig2} c) we display the situation for $\beta=1.75$ where the separation is 
larger but this latter ISS looks approximately the same as in the previous case, just slightly more curved and 
displaced towards a larger $k$ range. Finally, in Fig. \ref{fig2} d) new interactions with other 
ISSs become evident giving rise to what we will call other {\em coalescence-separation} smaller events, 
a phenomenon that can only be present for parameter spaces of dimension higher than 2.

\begin{figure}[t]
 \includegraphics[width=0.47\textwidth]{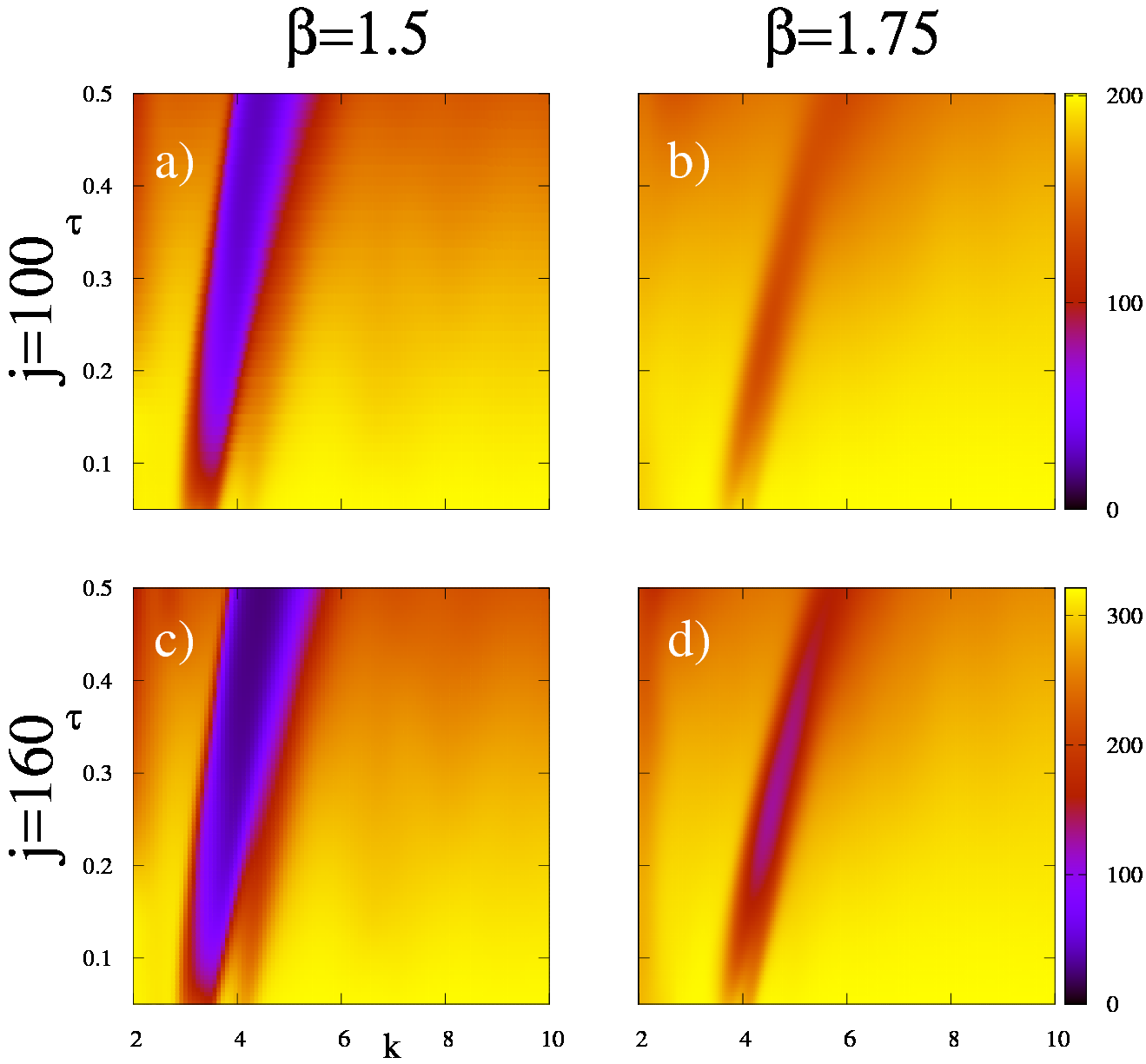}
\caption{(Color online) Quantum participation ratio $\eta$ in parameter space $(k,\tau)$ 
corresponding to $\beta=1.5$ and $\beta=1.75$, for $j=100$ (top left (a)) and right (b)) panels), 
and for $j=160$ (bottom left (c)) and right (d)) panels).} 
 \label{fig3}
\end{figure}

Now, we turn to analyze what is the quantum counterpart of this dynamics in the parameter space. 
For that purpose we select the cases $\beta=1.5$ and $\beta=1.75$ for which the separation is small 
and well developed, respectively. We have calculated the $\eta$ landscape for two different 
values of $j$ in order to also show the dependence on its size, an indicator of the semiclassical behaviour. 
Comparing Fig. \ref{fig3} a) 
for $\beta=1.5$ with Fig. \ref{fig3} b) for $\beta=1.75$ (both for $j=100$) we can see that the qISS 
reproduces the regular behaviour much better in the first case, though there is no significant 
difference at the classical level. This different behaviour persists as we go to the semiclassical 
limit, as is evident from Figs. \ref{fig3} c) and d). However, the overall quantum to classical agreement is 
better as expected. Then, why is there such a striking difference between the quantum behaviour at 
these two $\beta$ values for which the classical ISS is essentially the same? In the following we 
will answer this question by using some tools of our previously developed theory for quantum 
2D dissipative systems \cite{Carlo3,Carlo4}. 

\begin{figure}[t]
 \includegraphics[width=0.47\textwidth]{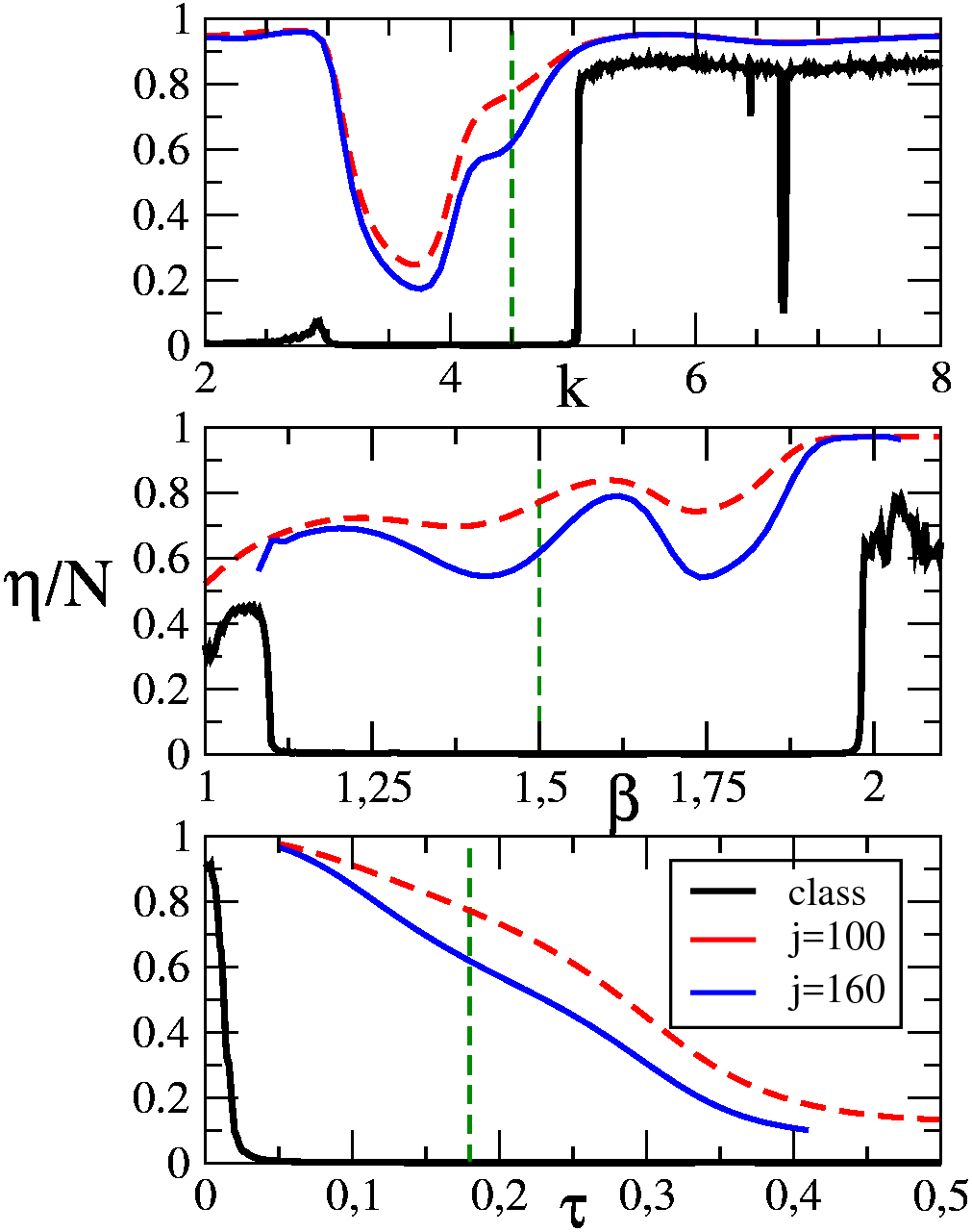}
\caption{(Color online) Normalized participation ratio $\eta/N$, with $N=1000$ for the classical case 
(black lines), and $N=2j+1$ for the quantum ones ($j=100$ ((red) gray dashed lines) 
and $j=160$ ((blue) gray lines)), 
as a function of $k$ (top panel), $\beta$ (middle panel), and $\tau$ (bottom panel). 
Parameter other than the axis one, takes the fixed value $k=4.5$, $\beta=1.5$ or $\tau=0.18$ 
(dashed (green) gray vertical lines).}
 \label{fig4}
\end{figure}

We first determine if the separation of the ISS from the main regular region 
is actually there at $\beta=1.5$. For that we show the quantum and classical 
normalized participation ratio 
along three lines in the direction of the axes of the parameter 
space, intersecting at $(k,\beta,\tau)=(4.5,1.5,0.18)$. In the top panel of Fig. \ref{fig4} we see 
that at approximately $k=3$ there is a small (but significant in terms of regular structures) 
rise of the classical participation ratio that clearly 
signals the separation of the ISS from the main regular region. At the quantum level just the 
ISS is resolved and there is some internal structure also. In the middle panel 
we notice that although the quantum participation ratio is generally lower inside the boundaries of the 
classical ISS the localization is not as strong as for the approximate interval $k \in [3.5;4]$ (see top panel). 
However some internal features are also present, in agreement with the previous results. 
Finally, in the bottom panel we see that localization monotonously increases as a function of $\tau$, 
which is something to be expected given the greater dissipation. 
In all cases the differences due to the size of $j$ are negligible indicating an extremely 
slow convergence to the classical limit without any further ingredients \cite{Carlo3}.

\begin{figure}[t!]
 \includegraphics[width=0.47\textwidth]{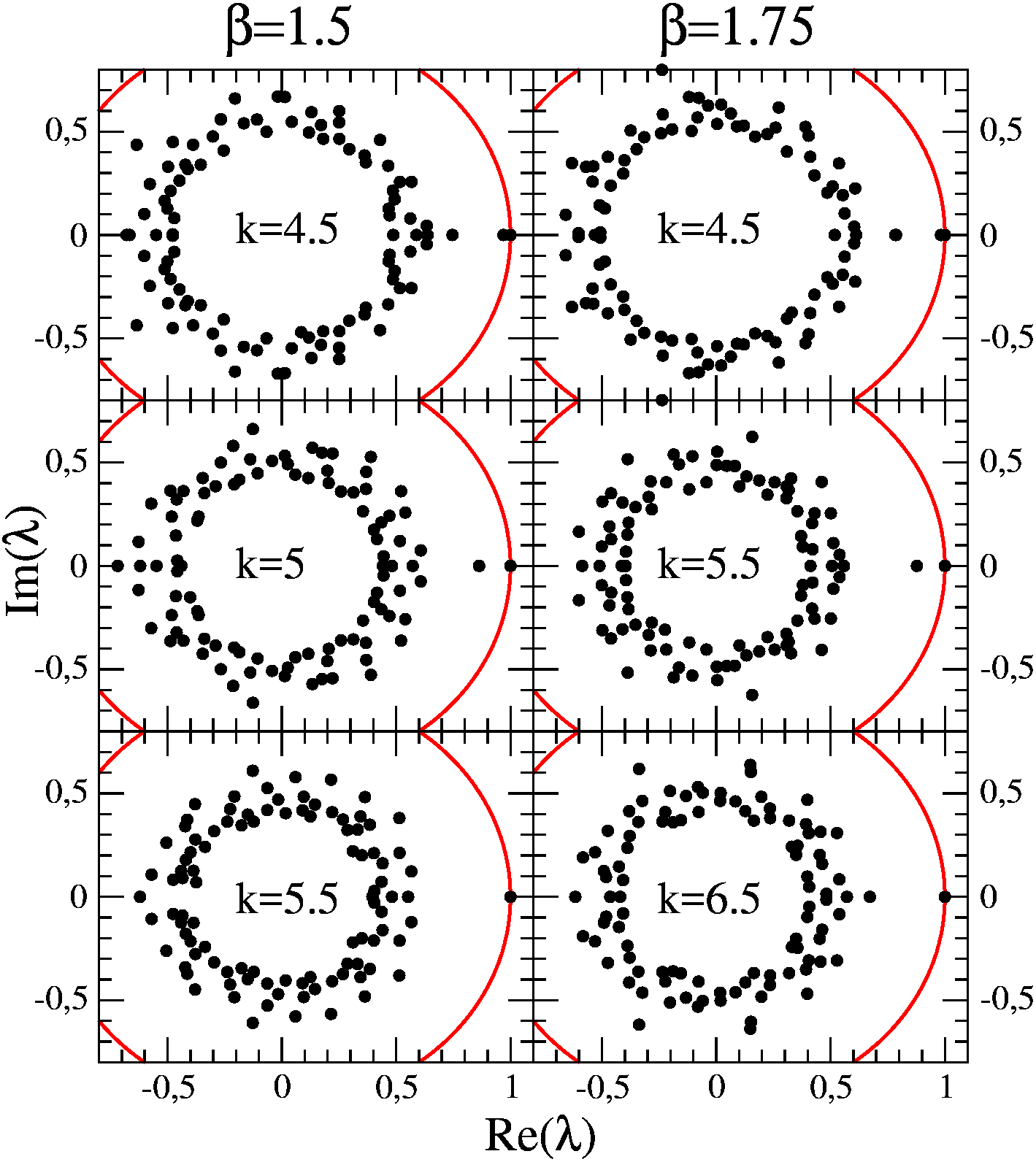}
\caption{(Color online) Quantum superoperator spectra in the complex plane.
Eigenvalues with largest moduli for $\beta=1.5$ and $\beta=1.75$ are shown in the left and right 
columns, respectively. From top to bottom $k=4.5$, $k=5$, and $k=5.5$ on the left; $k=4.5$, $k=5.5$, 
and $k=6.5$ on the right. We fix $j=160$ and $\tau=0.18$ in all cases.}
 \label{fig5}
\end{figure}

We now characterize the qISS at these two $\beta$ values by comparing the spectral and eigenstates 
behaviour at them. It is worth noticing that for the eigenstates representation we use the 
Wigner function on the sphere.
For a system of total angular momentum $j$, the density matrix $\hat{\rho}$ can be expressed in the 
Dicke representation as $\rho_{mm'}= \langle jm|\hat{\rho}|jm' \rangle$. Alternatively, we can 
consider the coupled total angular momentum representation, where 
\begin{equation}
\rho_{kq} =  \sum_{m=-j}^{j}\sum_{m'=-j}^{j} \rho_{mm'} t_{kq}^{jmm'},
\end{equation}
with 
\begin{equation}
\rho_{mm'}= \langle jm|\hat{\rho}|jm' \rangle = \sum_{k=0}^{2j}\sum_{q=-k}^{k} \rho_{kq}t_{kq}^{jmm'},
\end{equation}
and the Clebsch-Gordan transformation coefficients \cite{Dowl} given by 
\begin{equation}
t_{kq}^{jmm'} =(-1)^{j-m-q} \langle j,m;j,−m'|k,q \rangle.
\end{equation}
These latter are nonzero only if $q=m-m'$. Both representations contain the same information and 
are completely interchangeable. While the Dicke representation is more common, the coupled total angular 
momentum representation allows expressing the Wigner function on the Bloch sphere.
The Wigner function \cite{Schl} is a function on a sphere of radius $\sqrt{j(j+1)}$, 
represented in terms of orthonormal Laplace spherical harmonics as \cite{Dowl}
\begin{equation}
W(\theta,\phi)= \sum_{k=0}^{2j}\sum_{q=-k}^{k} \rho_{kq}Y_{kq}(\theta,\phi),
\end{equation}
where $\theta$ is the polar angle measured from the z axis, and $\phi$ is the azimuthal angle around 
the z-axis. This Wigner function contains the same information as the density matrix for any spin-j system. 
The marginals of the spherical Wigner function are the projection quantum number distributions along all 
quantization axes \cite{Schm}. In the following we will take the previously mentioned 
rescaled variable $\mu =\cos \theta$ instead of just $\theta$.

\begin{figure}[t]
 \includegraphics[width=0.47\textwidth]{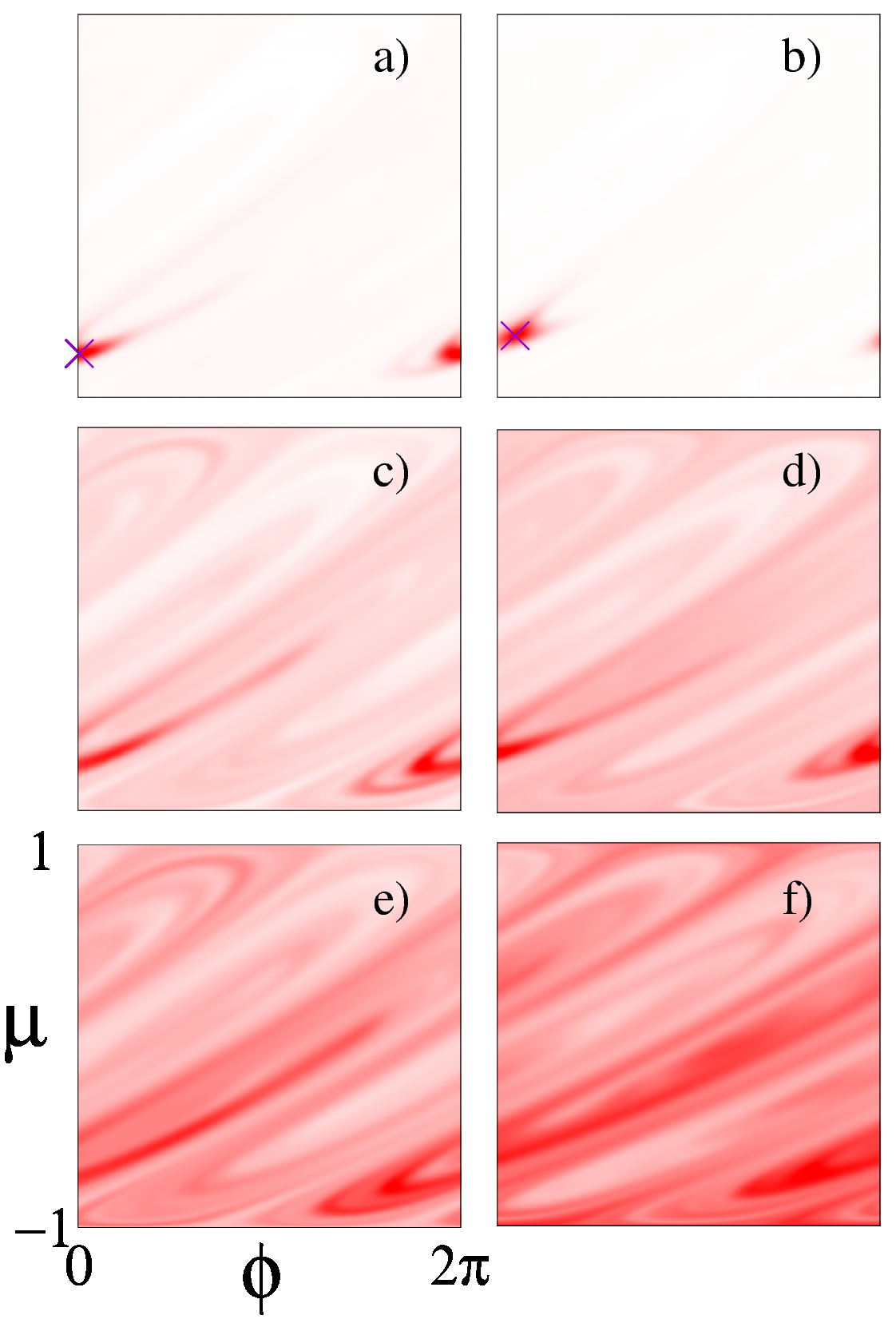}
\caption{(Color online) Wigner function of the invariant eigenstates, corresponding to eigenvalue $\lambda_1=1$ 
with the same parameter values of Fig. \ref{fig5}. $k=4.5$ (a), $k=5$ (c) and $k=5.5$ (e) for $\beta=1.5$, 
and $k=4.5$ (b), $k=5.5$ (d) and $k=6.5$ (f) for $\beta=1.75$. Classical corresponding limit cycles are marked 
with crosses.}
 \label{fig6}
\end{figure}

\begin{figure}[t]
 \includegraphics[width=0.47\textwidth]{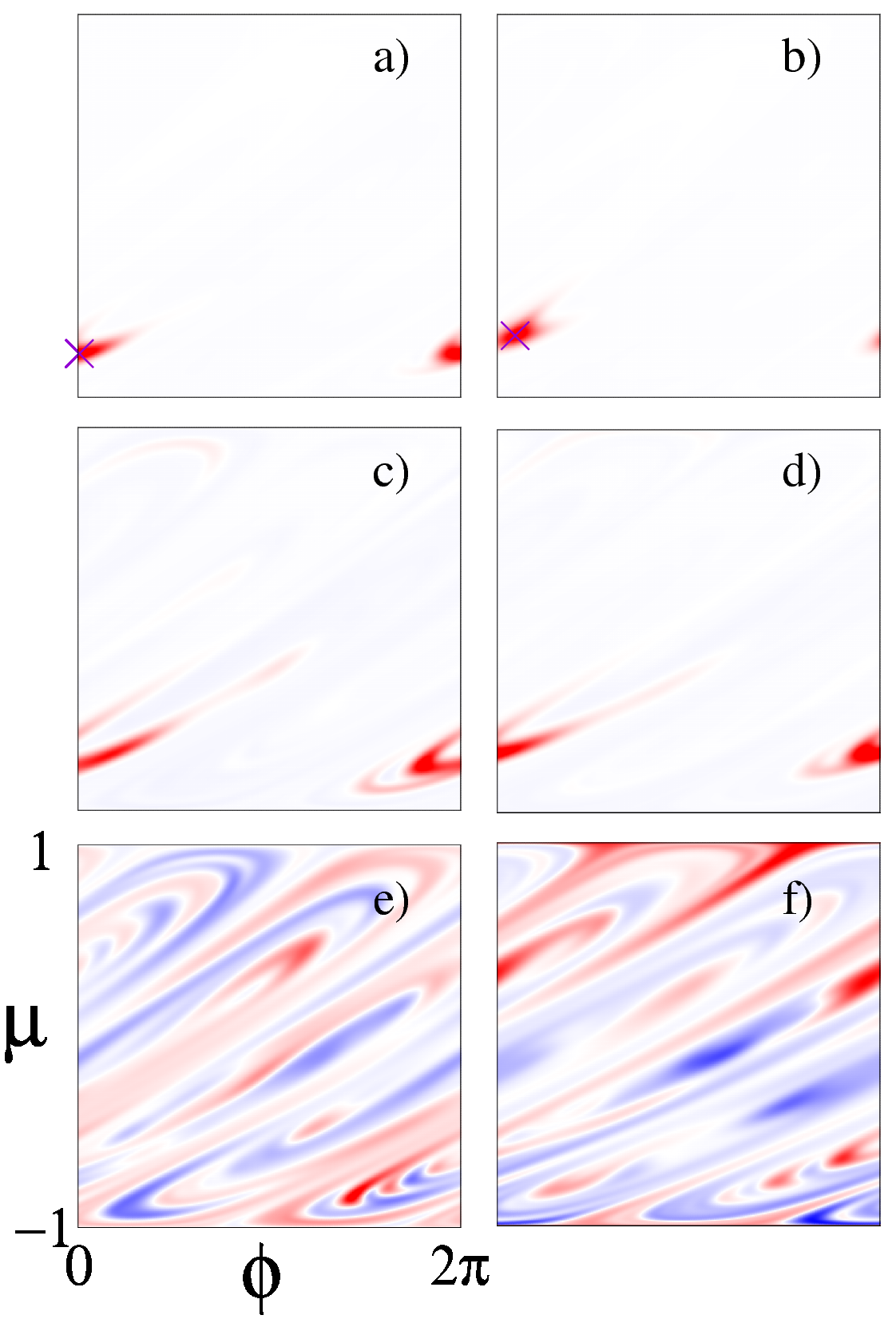}
\caption{(Color online) Wigner function of the leading eigenstates, corresponding to second largest 
eigenvalue $\vert\lambda_2\vert$ with the same parameter values of Fig. \ref{fig5}. 
$k=4.5$ (a), $k=5$ (c) and $k=5.5$ (e) for $\beta=1.5$, 
and $k=4.5$ (b), $k=5.5$ (d) and $k=6.5$ (f) for $\beta=1.75$. Classical corresponding limit cycles are marked 
with crosses.}
 \label{fig7}
\end{figure}

The spectra are displayed in Fig. \ref{fig5}, in the left column for $\beta=1.5$, 
and in the right one for $\beta=1.75$. 
Three different $k$ values have been considered in each case (always with $j=160$ and $\tau=0.18$). 
The corresponding invariant and leading eigenstates are shown in Figs. \ref{fig6} and \ref{fig7}, 
respectively (again, the left columns correspond to the lower $\beta$ value, and the rows respect the 
$k$ ordering of Fig. \ref{fig5}). 
The top panels of Fig. \ref{fig5} both show a leading real eigenvalue extremely close to the invariant 
one. This is a typical feature of large qISS whose invariant and leading eigenstates are localized around 
the corresponding classical limit cycle, within quantum uncertainty \cite{Carlo4}. 
This is clearly noticed by looking at Figs. \ref{fig6} and \ref{fig7} top panels. 
With crosses we mark the classical period 1 orbit that 
characterizes the large ISS and dominates the dynamics in this region of the parameter space. 
Moreover, when we move to larger $k$ values, but stay around the borders of the ISS the behaviour changes  
and the invariant eigenstates become chaotic for both $\beta$ values, 
as shown in the middle panels of Fig. \ref{fig6}. 
However, the leading eigenstates are localized around the same region of phase space corresponding 
to the limit cycle that belongs to the ISS, as can be seen in the middle panels of Fig. \ref{fig7}. 
The leading eigenvalues have moduli of $0.85$ approximately, indicating a still long decay towards 
the invariant. 
 Finally, when moving farther away from the ISS we find chaotic invariant and leading eigenstates 
as can be verified by inspecting the bottom panels of Figs. \ref{fig6} and \ref{fig7}. It is 
worth noticing that the Wigner function of the leading eigenstates in this case has real and imaginary 
part (we have displayed the real part) with positive and negative regions (red and blue colors respectively). 
The corresponding eigenvalues are no more real and the spectral gap is large. 
This is a generic behaviour similar to what we have found in the 2D case \cite{Carlo4}, 
and most importantly, it is a clear indication that the morphology of the qISS is the same for both $\beta$.

\begin{figure}[t!]
 \includegraphics[width=0.47\textwidth]{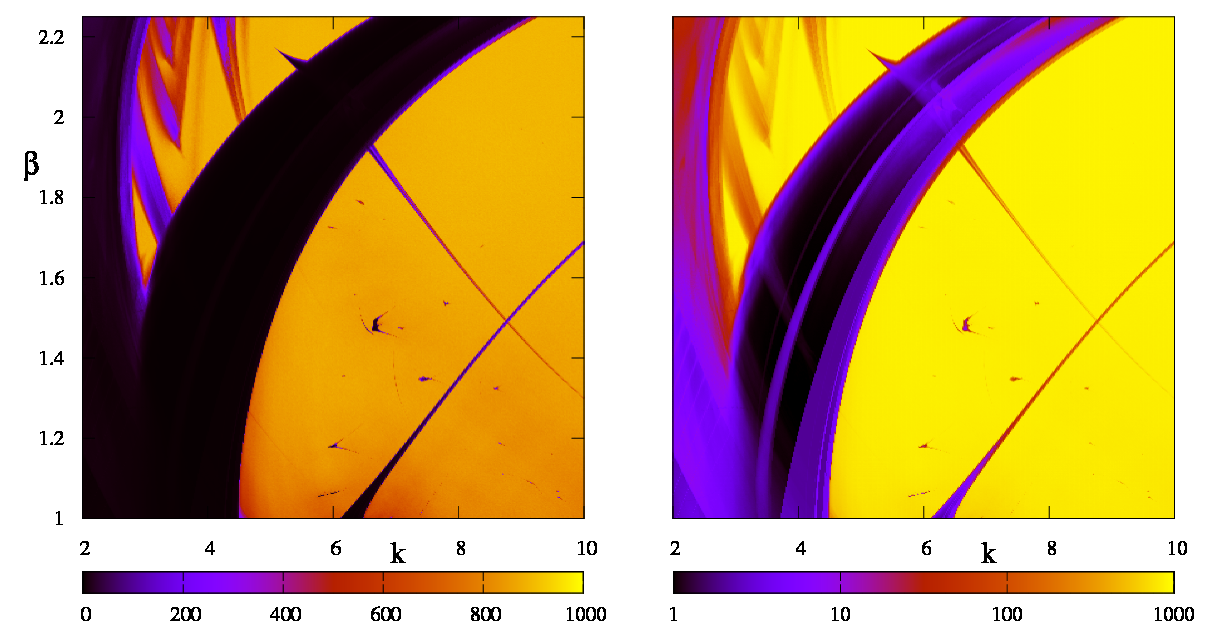}
\caption{(Color online) Participation ratio in parameter space $(k,\beta)$ 
for $\tau=0.18$ with linear and logarithmic (color) gray distribution in the left and 
right panels respectively.}
 \label{fig8}
\end{figure}

Then, how can we explain the marked quantum localization enhancement found for $\beta=1.5$ with respect to 
$\beta=1.75$ in Fig. \ref{fig3}? The explanation 
lies at the 3D nature of the parameter space. For spaces with more than 2 parameters the 
coalescence-separation phenomenon can take place, and quantum mechanically this could induce an 
enhancement of the region of localization. This can be better appreciated by means of the left panel of 
Fig. \ref{fig8} which shows a cut of the parameter space in the plane $\tau=0.18$. 
From approximately $\beta=1.5$ to lower values the ISS merges with the big regular region. 
This coalescence has no intermittencies and generates a very big regular area that we know from 
our previous studies can be better reflected in the quantum realm \cite{Carlo3}. In the right 
panel the logarithmic scale reveals an internal classical structure than explains the local minima 
found in the quantum participation ratio of Fig. \ref{fig4}.

\section{Conclusions}
\label{sec4}

We study the dissipative kicked top, a paradigmatic system in quantum and 
classical chaos which is also the starting point of many body models. It has 
a spherical phase space and a 3D parameter space. As a result we have extended the 
validity of localization properties of qISSs to this case, giving them a more 
generic nature. In fact, we identify the same effects of parametric tunneling found 
in 2 parameter systems that induces a chaotic shape for the invariant eigenstates 
in parameter regions corresponding to an ISS but that are near the chaotic background. 
Also, we have verified the localization on the limit cycles of the ISS for the 
leading eigenstates \cite{Carlo4}. 

On the other hand, we have found deep consequences of the {\em coalescence-separation} 
phenomenon, only present for systems with a parameter space with more than two dimensions. 
In fact, 3D ISSs can merge and break up as one of the parameters varies. Indeed, the quantum 
manifestations of this dynamics can be very important, leading to a marked enhancement of 
localization due to the enlargement of the regular regions. This can have very important 
derivations in many body dissipative systems where parameters proliferate. Even for the 
mean-field approximation one can have several of them \cite{Gambetta}. 

In the future, we will study the generalization to more dimensions. The first step would 
be to characterize the measure of this phenomenon at the classical level, which has not been 
addressed in the literature to the best of our knowledge. The next one would be to analyze 
the quantum counterparts and direct application to many body problems.

\section*{Acknowledgments}

One of us (LE) acknowledges support from ANPCYT under project PICT 2243-(2014).

\vspace{3pc}


\end{document}